\documentclass{elsarticle}

\AtBeginDocument{%
  \providecommand\BibTeX{{%
    \normalfont B\kern-0.5em{\scshape i\kern-0.25em b}\kern-0.8em\TeX}}}

\usepackage{booktabs}
\usepackage{ulem}
\usepackage{caption}
\usepackage{subcaption}
\usepackage{lineno,hyperref}
\modulolinenumbers[5]
\usepackage{rotating}
\usepackage{enumitem}

\usepackage{multirow}
\usepackage{graphicx}
\usepackage{balance}
\usepackage[linesnumbered,boxruled]{algorithm2e}
\usepackage{algorithm2e}
\usepackage{comment}
\usepackage{array}
\usepackage{algorithmic}

\usepackage{xpatch,environ,ragged2e}
\usepackage[colorinlistoftodos]{todonotes}
\RenewEnviron{abstract}{%
  \xdef\theabstracttext{%
    \unexpanded{%
      \def\baselinestretch{1}\noindent\unskip\textbf{Abstract}\par\medskip
      \noindent\unskip\ignorespaces}%
    \unexpanded\expandafter{\BODY}%
  }%
}
\def\theabstracttext{}
\xpatchcmd{\pprintMaketitle}
  {\ifvoid\absbox}
  {\ifx\theabstracttext\empty\else\printtheabstracttext\fi\ifvoid\absbox}
  {}{}
\newcommand{\printtheabstracttext}{{%
  \begin{trivlist}
  \normalfont\normalsize
  \item\relax
  \theabstracttext
  \end{trivlist}
}}

\usepackage{pgfplotstable}

\def\BibTeX{{\rm B\kern-.05em{\sc i\kern-.025em b}\kern-.08em
    T\kern-.1667em\lower.7ex\hbox{E}\kern-.125emX}}
\graphicspath{{pics/}}
\newcommand{\Cross}{cross-ecosystem library}
\newcommand{\Crosses}{cross-ecosystem libraries}

\newcommand{\RqOne}{\textbf{(RQ1)} \emph{What percent of contributors to a \Cross~repository are from different ecosystems?}}
\newcommand{\RqTwo}{\textbf{(RQ2)} \emph{How diverse are the programming languages for a \Cross~repository?}}
\newcommand{\PQ}{\textbf{(P1)} \emph{How dependent is the ecosystem to a \Cross~release?}}
\newcommand{\RqOneR}{
A significant majority (median of 37.5\%) of contributors come from a single ecosystem instead of from both ecosystems.
Interestingly, we find there is a significant portion of contributors do not belong
to any ecosystem (median of 24.06\%).
}
\newcommand{\RqTwoR}{
Three (PyPI, CRAN, RubyGems) out of the five ecosystems show the majority of the source code is written using languages that are not specific to an ecosystem (i.e., Python, Java, JavaScript, R, and Ruby).
}
\definecolor{chestnut}{rgb}{0.8, 0.36, 0.36}

\usepackage[most]{tcolorbox}

\tcbset{textmarker/.style={
        enhanced,
        parbox=false,boxrule=0mm,boxsep=0mm,arc=0mm,
        outer arc=0mm,left=5mm,right=3mm,top=7pt,bottom=7pt,
        toptitle=1mm,bottomtitle=1mm,oversize}}
        
\newtcolorbox{hintBox}{textmarker,
    borderline west={6pt}{0pt}{yellow},
    colback=yellow!10!white}
\newtcolorbox{importantBox}{textmarker,
    borderline west={6pt}{0pt}{red},
    colback=red!10!white}
\newtcolorbox{noteBox}{textmarker,
    borderline west={8pt}{0pt}{gray},
    colback=gray!10!white}

\newcommand{\summary}[1]{\begin{noteBox} \textbf{Summary:} #1 \end{noteBox}}

\begin{document}
\begin{sloppy}

\newcommand\rev[3]{\textcolor{red}{\sout{#1}} {\textcolor{blue}{#2}} {\todo[color=green!40]{\thesubsection{}. #3}}}

%\todo[size=\small, color=green!40]{#3}
%\listoftodos

\title{Intertwining Ecosystems: A Large Scale Empirical Study of Libraries that Cross Software Ecosystems}

%\author{Anonymous Authors}
\author[]{Kanchanok Kannee}
\ead{kanchanok.kannee.kg8@is.naist.jp}
\author[]{Supatsara Wattanakriengkrai}
\ead{wattanakri.supatsara.ws3@is.naist.jp}
\author[]{Ruksit Rojpaisarnkit}
\ead{rojpaisarnkit.ruksit.rn1@is.naist.jp}
\author[]{Raula Gaikovina Kula}
\ead{raula-k@is.naist.jp}
\author[]{Kenichi Matsumoto}
\ead{matumoto@is.naist.jp}

\address{Nara Institute of Science and Technology, Japan}

\begin{abstract}
An increase in diverse technology stacks and third-party library usage has led developers to inevitably switch technologies.
To assist these developers, maintainers have started to release their libraries to multiple technologies, i.e., a \Cross.
Our goal is to explore the extent to which these cross-ecosystem libraries are intertwined between ecosystems.
We perform a large-scale empirical study of 1.1 million libraries from five different software ecosystems, i.e., PyPI for Python, CRAN for R, Maven for Java, RubyGems for Ruby, and NPM for JavaScript to identify 4,146 GitHub projects that release libraries to these five ecosystems.
Analyzing their contributions, we first find that a significant majority (median of 37.5\%) of contributors of these cross-ecosystem libraries come from a single ecosystem, while also receiving a significant portion of contributions (median of 24.06\%) from outside their target ecosystems.
We also find that a \Cross~is written using multiple programming languages.
Specifically,  three (i.e., PyPI, CRAN, RubyGems) out of the five ecosystems has the majority of source code is written using languages not specific to that ecosystem.
As ecosystems become intertwined, this opens up new avenues for research, such as whether or not cross-ecosystem libraries will solve the search for replacement libraries, or how these libraries fit within each ecosystem just to name a few.
\end{abstract}

\maketitle
\section{Introduction and Motivation}
\label{sec:introduction}

Popular use of third-party libraries has become prominent in contemporary software engineering \citep{KulaEMSE2018}, which is evident by the emergence of different library repositories like NPM, PyPI, CRAN, Maven, and so on.
These massive repositories also depend on each other, thus forming a complex software ecosystem of dependencies, across these different technology stacks.

For various reasons, a developer may realize that a library used in their applications requires replacement.
One reason may be due to the availability of newer versions that fix defects, patch vulnerabilities, and enhance features.
In such cases, the developer seeks an appropriate replacement and has lead to various efforts in library recommendation \citep{Cossette2021recommendation}.
Another case is when a developer is forced to switch programming languages. 
With an increase of the diversity in technology stacks and third-party library usage \citep{Syful2021ICSME}, developers eventually will face the need to switch a programming language and their subsequent libraries specific to that language \citep{analogicalqa,Chen2016QA,Teyton2013map}.
In this case, developers would have to search for a replacement library that they are familiar with.

\begin{figure*}[]
    \centering
    \begin{subfigure}{1\linewidth}
        \includegraphics[width=.9\textwidth]{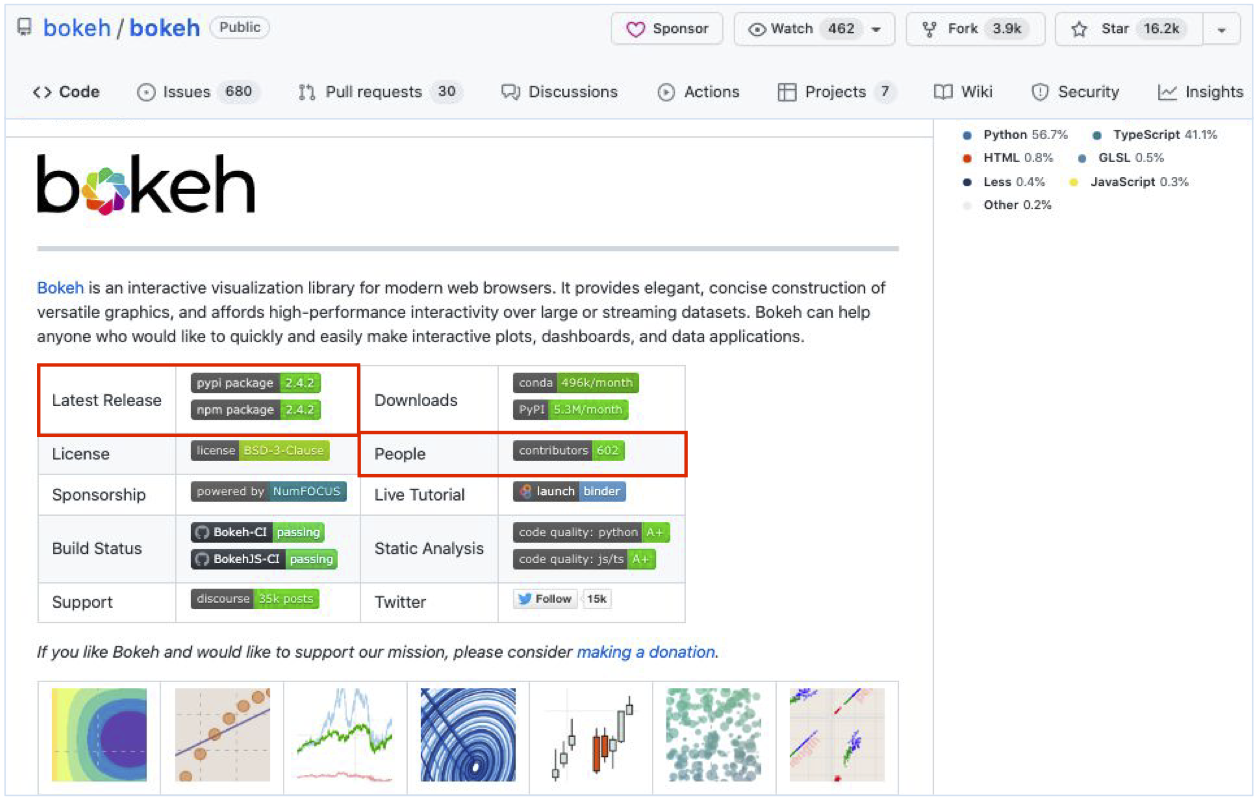}
        \caption{ The README file of the Bokeh project repository on that is hosted on GitHub}
    \end{subfigure}
    \begin{subfigure}{0.49\linewidth}
         \includegraphics[width=\textwidth]{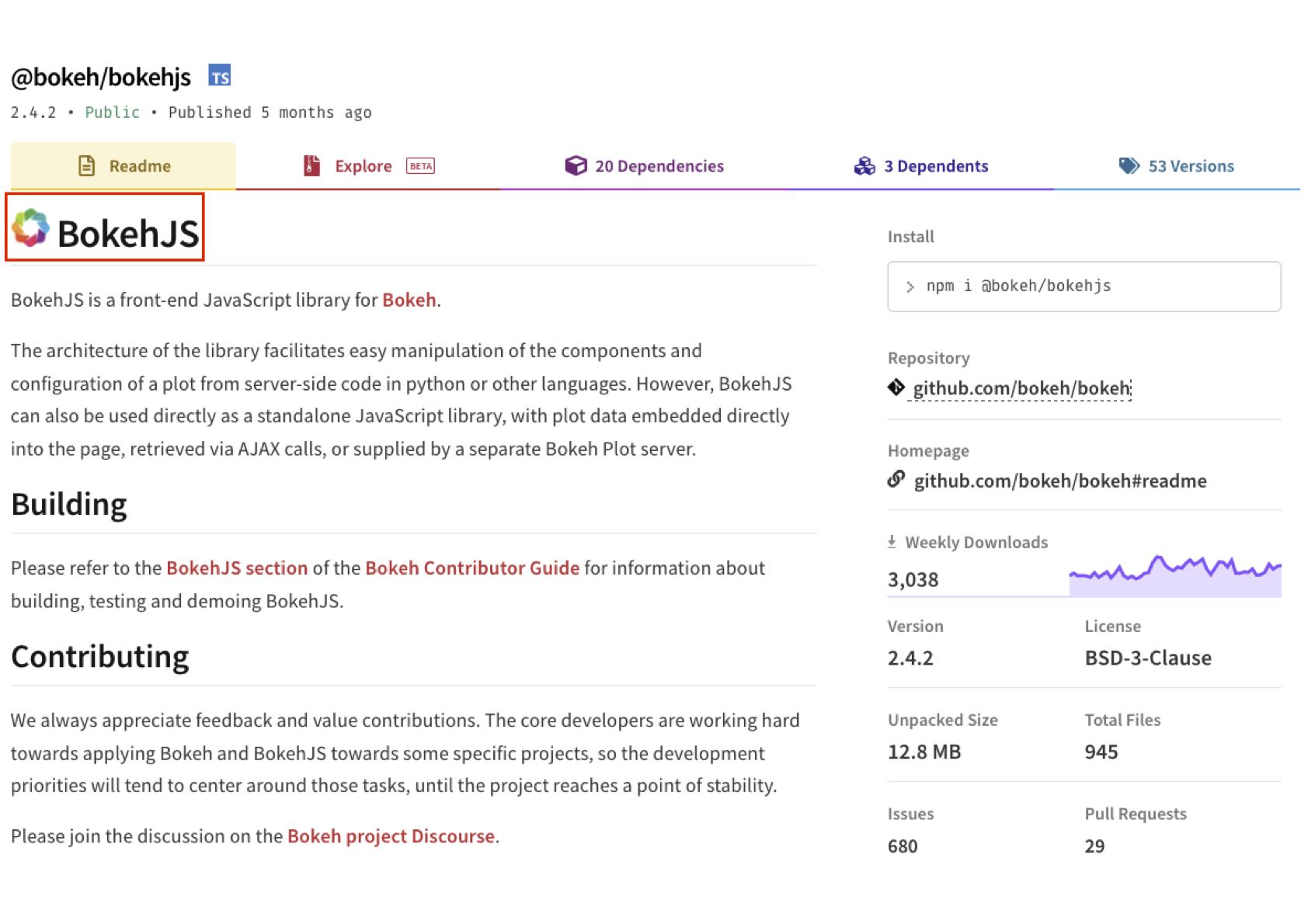}
         \caption{https://www.npmjs.com/package/@bokeh/bokehjs}
     \end{subfigure}
     \begin{subfigure}{0.49\linewidth}
         \includegraphics[width=\textwidth]{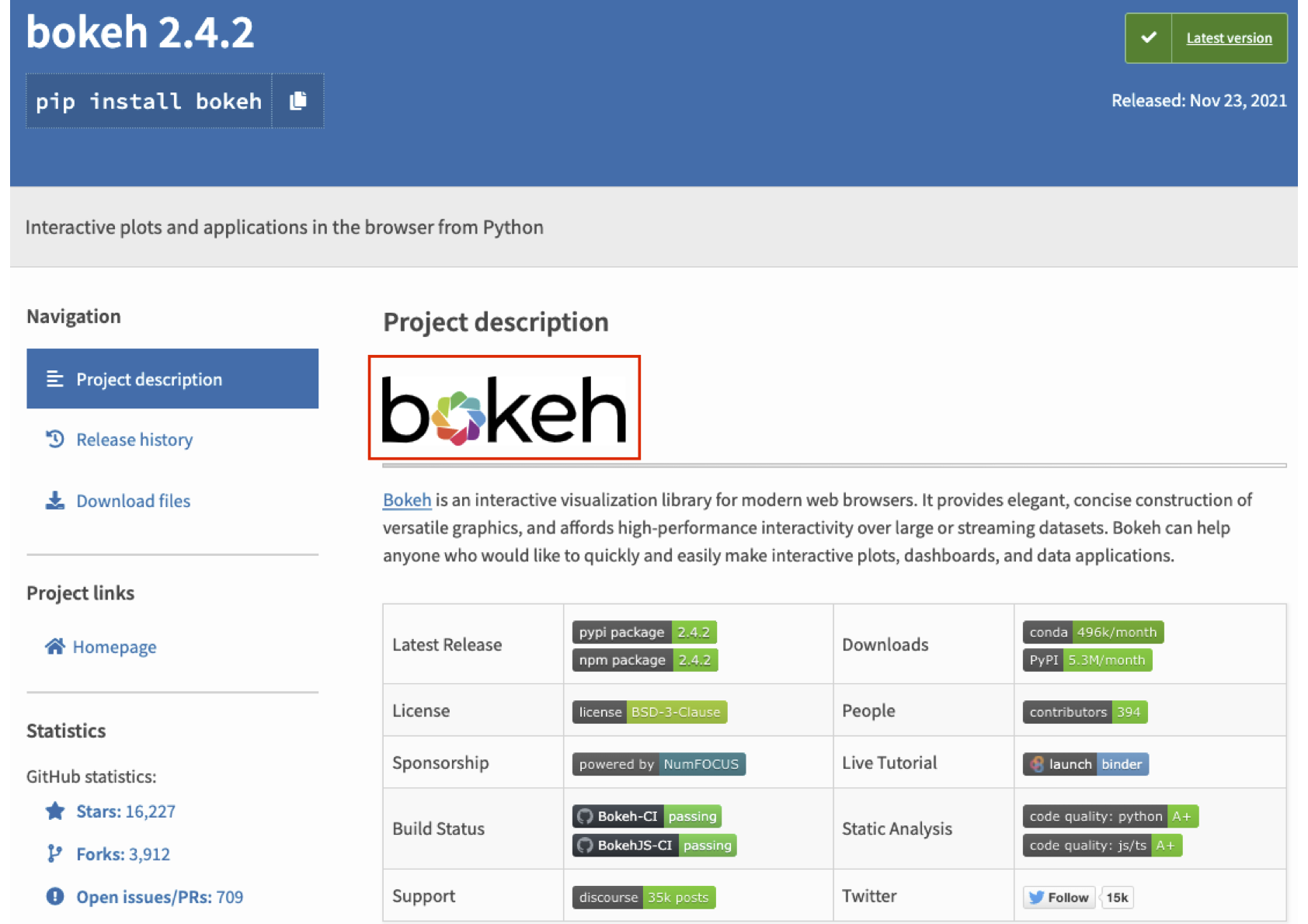}
         \caption{https://pypi.org/project/bokeh/}
     \end{subfigure}
    \caption{Our running example of Bokeh, a \Cross~that is released to the PyPI (Bokeh) and npm (BokehJS) ecosystems}
    \label{fig:bokeh}
\end{figure*}

In the case of switching technologies (i.e., programming languages), maintainers now provide an option by releasing a version specific for that ecosystem. We refer to this phenomenon as a \Cross.
Figure~\ref{fig:bokeh} depicts a cross-ecosystem library that serves two different ecosystems (cf. Figure~\ref{fig:bokeh}b and Figure~\ref{fig:bokeh}c).
The library is Bokeh\footnote{\url{https://github.com/bokeh/bokeh}},
which is a popular interactive visualization library for modern web browsers.
Figure \ref{fig:bokeh} shows the official GitHub repository that hosts the common repository, while there are two versions of the library hosted on both the official PyPI\footnote{\url{https://pypi.org/project/bokeh/}} and NPM\footnote{\url{https://www.npmjs.com/package/@bokeh/bokehjs}} registry.
In 2022, the Python release of Bokeh had over 2.94K dependent repositories, and 590 other libraries in the ecosystem that rely on this library.
It has 120 releases and was first released on October 25th, 2013.
The GitHub repository that hosts the \Cross~is mainly implemented in Python (i.e., 56.7\%) and 
TypeScript (i.e., 41.1\%).
According to its homepage, BokehJS is written primarily in TypeScript and not in JavaScript.
Furthermore, Bokeh attracts has 603 contributors to its GitHub repository.

Our goal is to explore the extent to which these \Crosses~are intertwined within these ecosystems.
By studying developer contributions and programming language diversity, we set out to investigate the extent to which cross-ecosystem libraries by mining five of the most popular and widely adopted library ecosystems (i.e., CRAN, Maven, PyPI, RubyGems, and NPM).
Our study is a large-scale quantitative analysis of 1,110,059 libraries to identify 4,146 \Cross~with 567,864 contributors.
After preliminary results showed that \Crosses~are being depended upon, we formulated two research questions that analyze \Cross~repository on GitHub:

\begin{itemize}
\item \noindent \RqOne~\\
\textit{Motivation:} 
Prior work \citep{fse2018_sustained} shows that the ecosystem plays an important role in the sustained activities of a software project. Hence, we would like to test the assumption that a \Crosses~may require involvement from a target ecosystem.
\\
\textit{Results:} \RqOneR
\item \noindent \RqTwo~\\
\textit{Motivation:} 
To complement RQ1, the motivation for RQ2 is to specifically analyze which programming languages are implemented in a \Cross.
As shown in the motivating example, some libraries are only front-end implementation.
Hence, we would like to understand the extent to which these libraries are implemented for the different programming languages that serve the ecosystem (e.g., NPM for JavaScript).
\\
\textit{Results:} \RqTwoR 
\end{itemize}

In addition to the insights revealed by the study, our contributions are two-fold.
The first contribution is a large quantitative study that covers over 1.1 million libraries, 0.5 million contributors from five different software ecosystems.
The second contribution is a replication package with all data and scripts that contains all the algorithms used in the study.
Our preliminary appendix is available at \url{https://zenodo.org/record/6983864#.YvVmGuxBy3I}.

%%%%%%%%%%%%%%%%%%%%%%%%%%%%%%%%%%%%%%%%%%%%%%%%%%
\begin{table}[]
\centering
\caption{Overview of dataset for the preliminary analysis.}
\label{tab:collected_repo}
\scalebox{1}{
\begin{tabular}{lcc}
\hline
\multicolumn{1}{c|}{\multirow{2}{*}{\# Lib. Releases}} & \multicolumn{2}{c}{As of 12 Jan 2020} \\ \cline{2-3} 
\multicolumn{1}{c|}{}                                             & with GitHub Repo URL    & initial libraries  \\ \hline
\multicolumn{1}{l|}{ NPM}                                & 818,787                 & 2,357,829       \\
\multicolumn{1}{l|}{ PyPI}                                    & 138,001                 & 420,350        \\
\multicolumn{1}{l|}{ CRAN}                                    & 5,551                   & 21,526       \\
\multicolumn{1}{l|}{ Maven}                                   & 36,762                  & 456,756   \\
\multicolumn{1}{l|}{ RubyGems}                                & 110,958                 & 176,987    \\ \hline
\multicolumn{1}{c|}{Total}                                & 1,110,023                & 3,433,448   \\
\multicolumn{1}{c|}{\# \Cross~pairs}                       & \multicolumn{2}{c}{4,146}   \\\hline
\end{tabular}
}
\end{table}
%%%%%%%%%%%%%%%%%%%%%%%%%%%%%%%%%%%%%%%%%%%%%%%%

%%%%%%%%%%%%%%%%%%%%%%%%%%%%%%%%%%%%%%%%%%%%%%%%
\begin{figure*}[h]
    \centering
    \frame{\includegraphics[width=9cm]{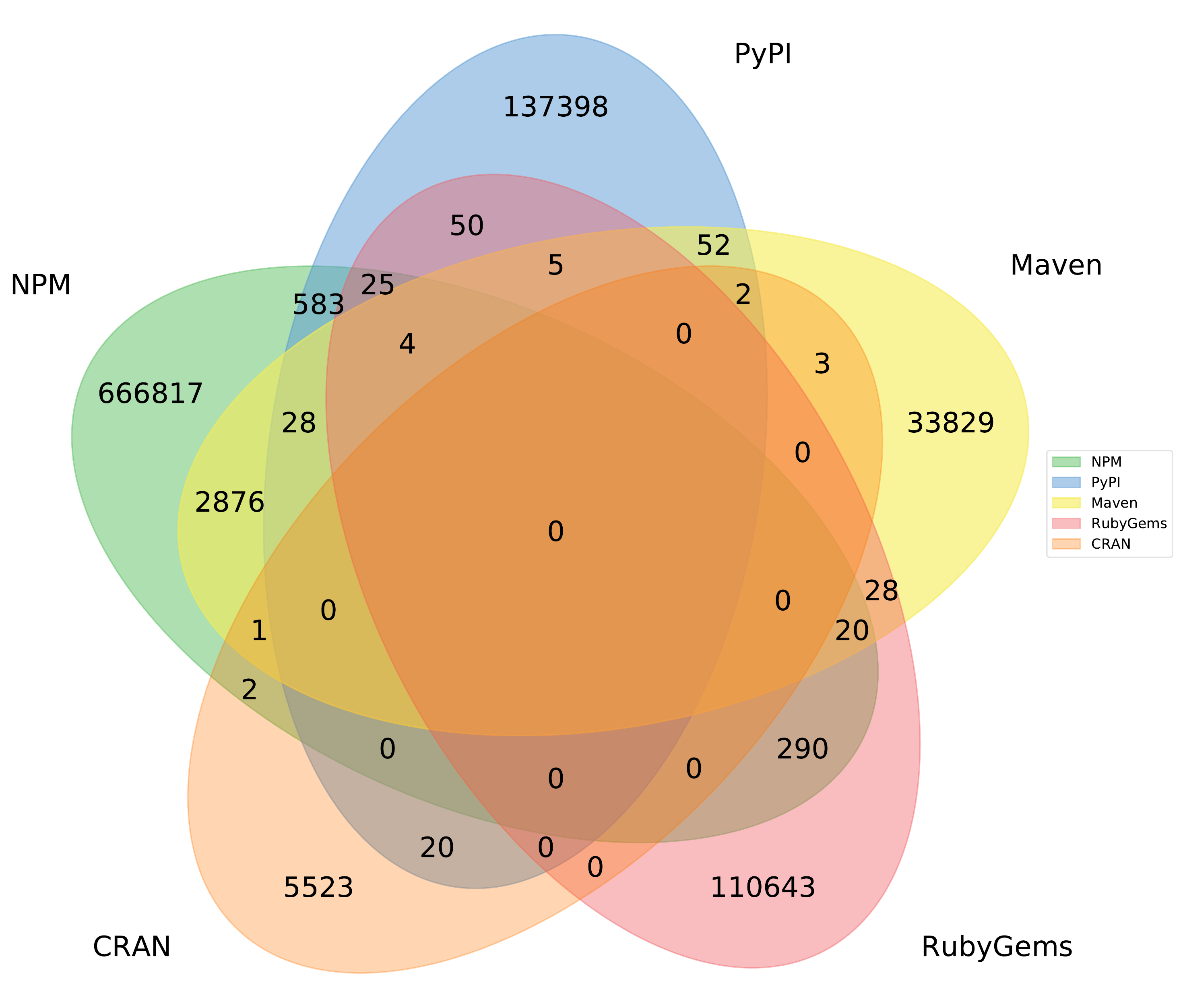}}
    \caption{Identifying the 4,146 cross-ecosystem libraries from the five ecosystems. For example, we identified 22 libraries across CRAN and PyPI.}
    \label{fig:RQ1_violinplot}
\end{figure*}
%%%%%%%%%%%%%%%%%%%%%%%%%%%%%%%%%%%%%%%%%%%%%%%%

\section{Preliminary Analysis}
To motivate the importance of a \Cross~on the ecosystem, we performed a preliminary study to answer the question \PQ

\paragraph{\textit{Data Collection}}
We selected five popular and well-studied software ecosystems. 
NPM is a package manager for the JavaScript programming language that was recently purchased by Microsoft via GitHub\footnote{\url{https://www.npmjs.com/}}.
PyPI is the library ecosystem that serves the Python programming language, which is interpreted high-level general-purpose programming language.
CRAN is the library ecosystem that serves the R programming language, which is a free software environment for statistical computing and graphics \footnote{\url{https://www.r-project.org/}}.
Maven is the library ecosystem that serves the Java programming language,  which is a general-purpose programming language that follows the object-oriented programming paradigm and can be used for desktop, web, mobile, and enterprise applications\footnote{\url{https://www.java.com/}}.
RubyGems is the library ecosystem that serves the Ruby programming language, which is a dynamic, open-source programming functional programming language with a focus on simplicity and productivity\footnote{\url{https://www.ruby-lang.org/en/}}.

To identify a \Cross, we use the GitHub library repository as the linking heuristic.
For instance, as shown in our motivating example, the GitHub library repository URL serves as a link between two analogical libraries.
Our assumption is that GitHub should be the common platform by which \Crosses~may decide to host their libraries.
We also use this filter as a quality check, to ensure that library quality.

Similar to prior work \citep{zerouali2019diversity,decan2019empirical,alfadel2021use}, we then queried the libraries.io dataset for library ecosystems that indeed listed a GitHub repository URL as their library repository.
For each library ecosystem, we collect a list of all libraries from the Libraries.io\footnote{\url{https://libraries.io/}} dataset.
We mined the dataset version (1.6.0)\footnote{\url{https://zenodo.org/record/3626071\#.YdRu_hNBzFo}}.
Once we were able to collect the list of libraries that hosted their library repository on GitHub, we then proceed to cross-reference between libraries that are hosted in different library ecosystems.
The results are shown in Table \ref{tab:collected_repo}, where we extracted a total of 1,110,023 from 3,433,448 libraries. 
As shown in Figure \ref{fig:RQ1_violinplot}, we noticed that the \Cross~is usually comprised in pairs. 
In other words, we identify a \Cross~as having releases to two ecosystems (e.g., NPM $\cap$ PyPI).

%%%%%%%%%%%%%%%%%%%%%%%%%%%%%%%%%%%%%%%%%%%%%%
\begin{table}[]
\caption{Summary statistics of \#Dependent showing statistical difference including significance where cross-ecosystem is more dependent than regular libraries.}
\label{tab:preliminary}
\begin{tabular}{cc|rcc|}
\cline{3-5}
\multicolumn{1}{l}{}                                                                     & \multicolumn{1}{l|}{} & \multicolumn{3}{c|}{{\# Dependent (regular)}}                                        \\ \cline{3-5} 
\multicolumn{1}{l}{}                                                                     & \multicolumn{1}{c|}{cross-ecosystem pair} & {Mean}                           & {Median}               & {Significance}      \\ \hline
\multicolumn{1}{|c}{}                                                                    & NPM $\cap$                  & 2664.71(761.39)                                      & 53(1)                            & * S               \\
\multicolumn{1}{|c}{\multirow{-2}{*}{}}       & PyPI                  & 405.35(29.43)                                       & 26(0)                             & - \\ \hline
\multicolumn{1}{|c}{}                                                                    & NPM $\cap$                  & 2884.94(761.39)                                       & 71(1)                             & -                     \\
\multicolumn{1}{|c}{\multirow{-2}{*}{}}      & Maven                 & 75.06(59.35)                                    & 5(0)                             & * N                    \\ \hline
\multicolumn{1}{|c}{}                                                                    & NPM  $\cap$                 & 4003.08(761.39)                                       & 19(1)                             & -                               \\
\multicolumn{1}{|c}{\multirow{-2}{*}{}}   & RubyGems              & 2291.94(335.90)                                       & 16(0)                             & -                               \\ \hline
\multicolumn{1}{|c}{}                                                                    & CRAN $\cap$                 & 21.43(19)       & 12(0) & -  \\
\multicolumn{1}{|c}{\multirow{-2}{*}{}}      & PyPI                  & -                                       & -                             & -                \\ \hline
\multicolumn{1}{|c}{}                                                                    & Maven  $\cap$               & 94.21(59.35)    & 20(0) &- \\
\multicolumn{1}{|c}{\multirow{-2}{*}{}}     & PyPI                  & 91.42(29.43)    & 23(0) & -   \\ \hline
\multicolumn{1}{|c}{}                                                                    & Maven  $\cap$               & 269.47(59.35)   & 34(0) & * S \\
\multicolumn{1}{|c}{\multirow{-2}{*}{}} & RubyGems              & 1104.91(335.90) & 56(0) &   * S \\ \hline
\multicolumn{1}{|c}{}                                                                    & PyPI   $\cap$               & 132.66(29.43)   & 7(0)  &  * S  \\
\multicolumn{1}{|c}{\multirow{-2}{*}{}}  & RubyGems              & 569.63(335.90)  & 8(0)  &  * S \\ \hline
\multicolumn{5}{l}{\begin{tabular}[c]{@{}l@{}}The effect sizes level: N(negligible), and S(small)\\ *:p-value 	$<$ 0.05\end{tabular}} 
\end{tabular}
\end{table}
%%%%%%%%%%%%%%%%%%%%%%%%%%%%%%%%%%%%%%%%%%%%%%

\paragraph{\textit{Approach}}
To answer this preliminary question, we collected dependent metrics, which is the number of other libraries that declare the library as a dependency.
Note that for the analysis, we compare against a pair of ecosystems (i.e., NPM $\cap$ PyPI).
For example, Font-Awesome\footnote{\url{https://github.com/fortawesome/Font-Awesome}} is one of the top open-source libraries on GitHub. the library has a dependency score for both the PyPI (i.e., 29 dependents) and Maven (i.e., 22 dependents).
We used the libraries.io dataset to append to our existing dataset.

For evaluation, we set up an experiment to compare the \Cross~against other regular libraries. 
As a result, we will report the statistical summary (i.e., mean, median, and standard deviation).
Furthermore, we statistically validate our results using the Mann-Whitney U test~\citep{mann1947test}, \citep{wilcoxon1945individual} which is a non-parametric statistical test.
To show the power of differences between metrics from cross-ecosystem libraries and regular libraries, we investigate the effect size using Cliff’s $\delta$, which is a non-parametric effect size measure \citep{romano2006exploring}. 
The interpretation of Cliff's $\delta$ is shown as follows: (1) $\delta$ $<$ 0.147 as Negligible,
(2) 0.147 $\leq$  $\delta$  $<$ 0.33 as Small, (3) 0.33 $\leq$  $\delta$  $<$ 0.474 as Medium, or (4)
0.474 $\leq$  $\delta$  as Large. To analyze Cliff’s $\delta$, we use the cliffsDelta package.
\footnote{\url{https://github.com/neilernst/cliffsDelta}}

\paragraph{\textit{Results}}
Table \ref{tab:preliminary}
shows the evidence that the cross-ecosystem library influences ecosystems.
We find that the cross-ecosystem library has more dependents than the regular library for every pair of cross-ecosystem libraries, e.g., the median of the number of dependents for the \Cross~that was released to the NPM and PyPI ecosystem: 53 dependent $<$ 1 dependents.
Statistically, we find that the number of dependents between cross-ecosystem and regular libraries are significantly different (p-value $<$ 0.05), with a negligible to small association as the reported effect size.
Furthermore, as shown in Table \ref{tab:preliminary}, we show that \Cross~combinations that include RubyGems, which is the pair of Maven with RubyGems and PyPI with RubyGems had a significantly larger number of dependents than regular libraries.

\summary{
We find that \Crosses~are relied upon more when compared to regular libraries, especially for Maven, RubyGems, and PyPI.
}

\section{Findings}
\label{sec:Findings}

Motivated by the preliminary results, which show evidence that cross-ecosystem libraries are relied upon by the ecosystem,
we now continue to answer each research question that relates to their contributions and diversity of programming language composition.

\begin{table}[]
\caption{Summary statistics of appended data to the 1,110,023 libraries to answer our RQs.}
\label{tab:data_for_each_rq}

\scalebox{0.95}{
\begin{tabular}{l|ccr}

\multicolumn{1}{l}{} &  As of            &  March 2022            &                \\
\cline{2-4}
                                               & {Median} & {Max} & {Total} \\ \hline
\multicolumn{1}{l|}{\textbf{RQ1 -\# Contrib.}} &                 &              &                \\
\multicolumn{1}{r|}{per cross-ecosystem lib.}                  & 8               & 1,727        & 49,674         \\
\multicolumn{1}{r|}{per lib.}     & 4               & 16,606       & 567,864        \\ \hline
\multicolumn{1}{l|}{\textbf{RQ2 -\# Lang. per lib.}}    & 2               & 44           & 120            \\ \hline
\end{tabular}}
\end{table}
%%%%%%%%%%%%%%%%%%%%%%%%%%%%%%%%%%%%%%%%%%%%%%%%

\subsection{Data Preparation}
\label{sec:data_collection}
As shown in Table \ref{tab:data_for_each_rq}, extending our dataset from the preliminary study, we mined and collected contributor information (RQ1) and programming language (RQ2).
It is important to note that RQ1 analyzes all cross-ecosystem libraries against regular libraries, while RQ2 only analyzes the cross-ecosystem libraries.

To answer RQ1 and RQ2, we use the GitHub API to collect all contributors that made commits to all repositories. 
We use the API request \texttt{https://api.github.com/repos/$\{$owner$\}$/$\{$repo$\}$/commits} to collect all the commit information.

After collecting all contributors for each library, we then merged the contributors' list based on the ecosystem, so that we have a merged listing of contributors.
For RQ2, we use the GitHub API to collect the lines of code that each programming language.
We use the API command \texttt{https://api.github.com/repos/$\{$owner$\}$/$\{$repo$\}$/languages} to retrieve all these informations.
This process was extremely time-consuming, taking almost two months to slowly collect all this information.

\subsection{Distribution of Contributors (RQ1)}
\begin{figure}[]
    \centering
    \frame{\includegraphics[width=6cm]{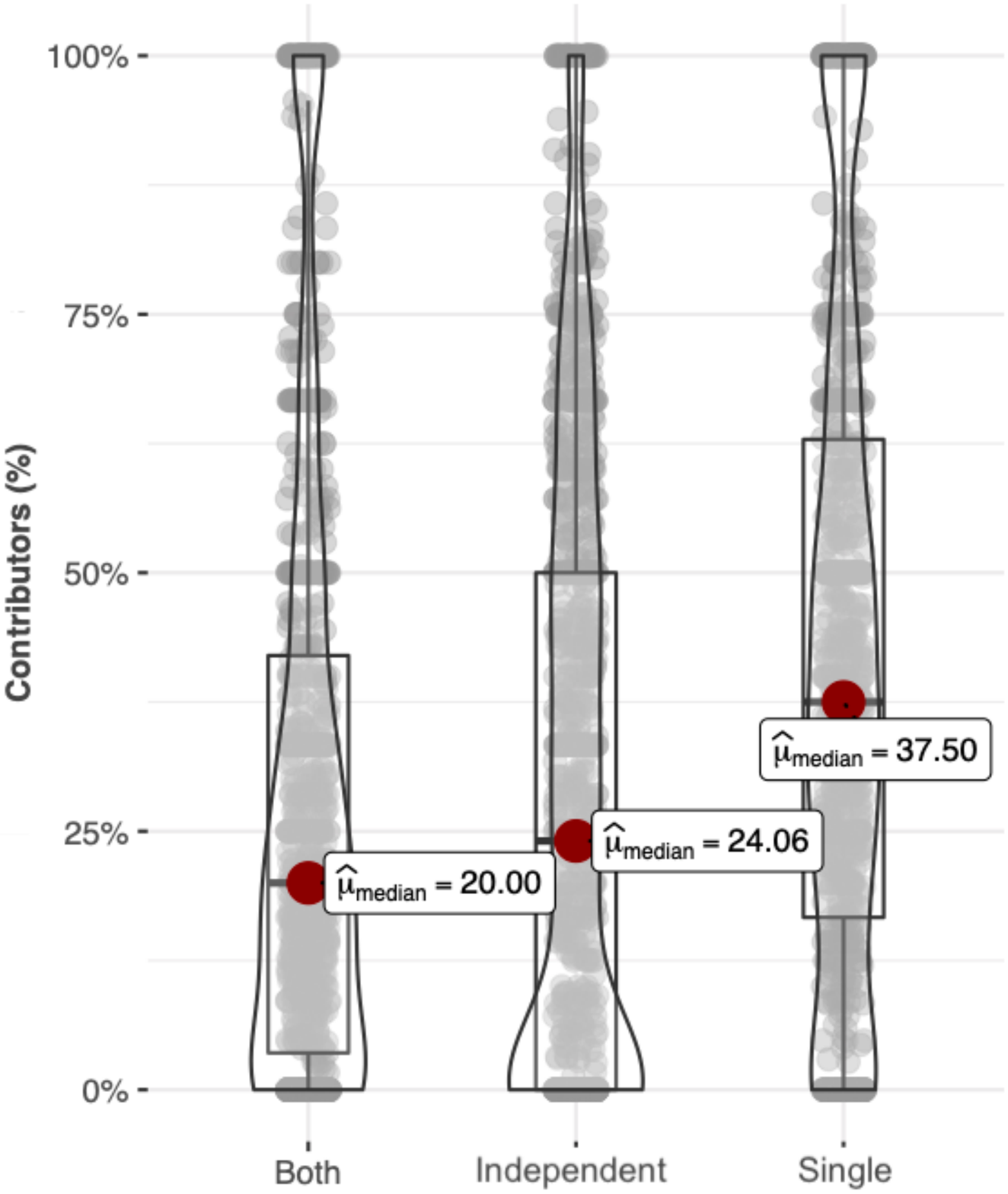}}
    \caption{Answering RQ1, we show how contributions for a \Cross~originate from a single ecosystem.}
    \label{fig:RQ1_violinplot}
\end{figure}

\paragraph{Approach}

To answer RQ1, we investigate whether a \Cross~attracts contributions from the targeted ecosystems.
Taking our example shown in Figure \ref{fig:bokeh}, we would like to understand the extent to which both the PyPI and NPM ecosystems support the GitHub Bokeh repository.
In terms of our motivating example, we calculate the percentage of the 533 contributors from the Bokeh repository has also made contributions to a regular PyPI or NPM repository.
For data collection, we first collected all contributors from the 1,110,059 projects.
For our analysis, we calculate a percentage of contributors from each \Cross~into these three classifications:
\begin{itemize}
    \item \textbf{Both (\%)} Percentage of contributors that have made prior contributions to libraries that belong to both ecosystems.
    \item \textbf{Single (\%)} Percentage of contributors that have made prior contributions to a single ecosystem.
    \item \textbf{Independent (\%)} Percentage of contributors do not have any contributions to any ecosystem libraries.
\end{itemize}
Since we are dealing with a large dataset, we used the PySpark\footnote{\url{https://spark.apache.org/docs/latest/api/python/}} arrays to classify and merge contributors into these three groups.
Hence, for a \Cross, we calculate these ratios with the hypothesis that most contributors would come from a single ecosystem.

To statistically confirm our classification of contributors, we use McNemar’s Chi-Square test \citep{McNemar1947}. This is a non-parametric statistical test to use to find the change in proportion for paired data and to find the change in proportion for our paired data. 
We test the null hypothesis that \textit{`the percentage of contributions classified is the same'}
We also measure the effect size using Cohen’s $d$, a non-parametric effect size measure by \citet{cohen2013statistical}. Effect size is analyzed as follows: 
(1) $ d < 0.2$ as Negligible, (2) $0.2 \leq d <0.5$ as Small, (3) $0.5 \leq d <0.8$ as Medium, or  (4) $0.8 \leq d$ as Large.

\paragraph{\textit{Most contributors originate from a single ecosystem}}

From Figure \ref{fig:RQ1_violinplot} we make two observations. 
First, we find that there is a higher percentage of contribution (a median of 37.5\%) that is supported by only one ecosystem. (see Figure~\ref{fig:RQ1_violinplot}).
It is important to note that our analysis does not identify the originating ecosystem.

The second observation is that a significant portion of contributors  (median of 24.06\%) that do not belong to either ecosystem.
The result shows that these contributors may be just specific to the project itself. 
This is more than contributors that belong to both ecosystems (median of 20\%).
The result indicates that these libraries may need to have their own base of contributors in order to sustain maintenance activities.

\begin{table}[]
\centering
\caption{Statistical significance test results related to RQ1. }
\label{tab:sta_RQ1}
\scalebox{0.8}{
\begin{tabular}{cc}
\hline
\multicolumn{2}{|c|}{\textbf{Contributions (\%)}}                                                                          \\ \hline
\multicolumn{1}{|c}{{Single-Both *}}                        & \multicolumn{1}{c|}{S}                     \\
\multicolumn{1}{|c}{{-}}                     & \multicolumn{1}{c|}{-}                     \\ 
\multicolumn{1}{|c}{Both-Neither *}                     & \multicolumn{1}{c|}{N}                     \\ \hline
\multicolumn{2}{l}{\begin{tabular}[c]{@{}l@{}}The effect sizes level: small(S) and negligible(N)\\ *:p-value \textless 0.05\end{tabular}}
\end{tabular}}
\end{table}

%%%%%%%%%%%%%%%%%%%%%%%%%%%%%%%%%%%%%%%%%%%%%%%%
\begin{figure*}[!]
    \centering
    \begin{subfigure}{0.3\linewidth}
        \includegraphics[width=\textwidth]{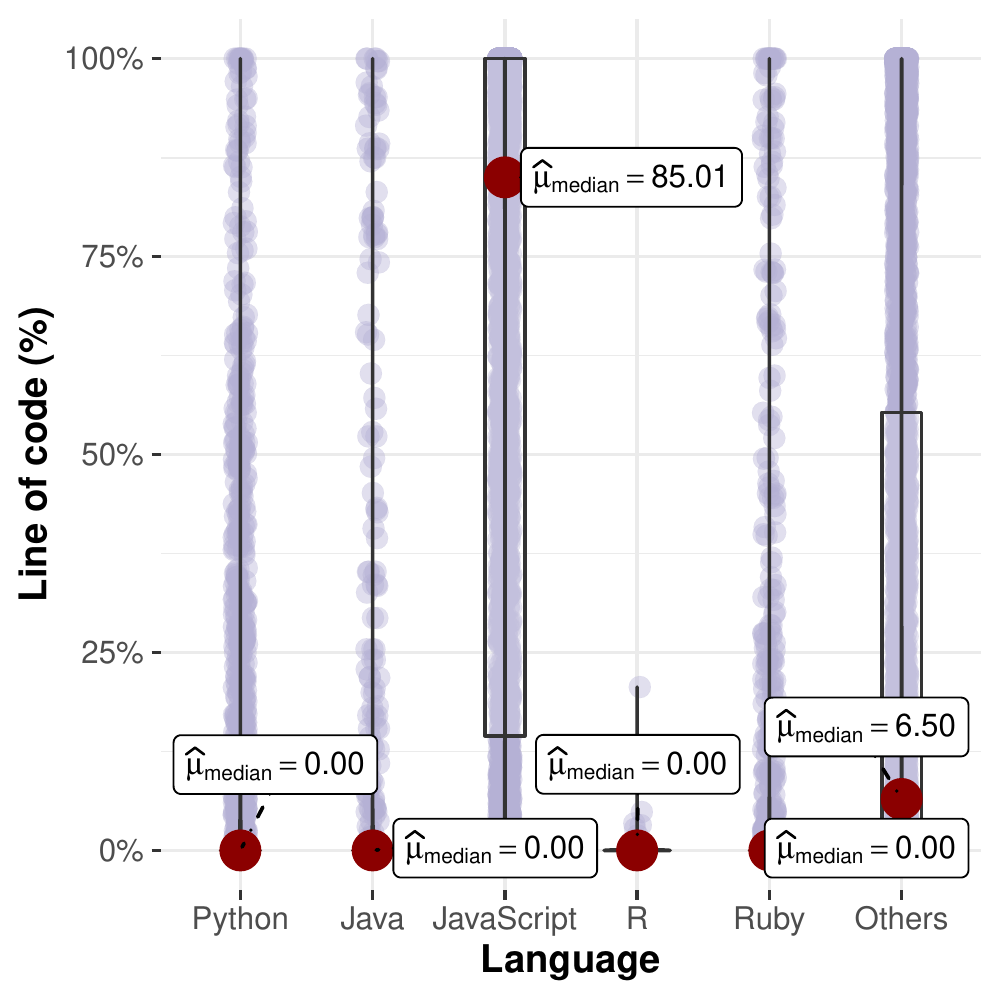}
        \caption{NPM}
    \end{subfigure}
    \begin{subfigure}{0.3\linewidth}
         \includegraphics[width=\textwidth]{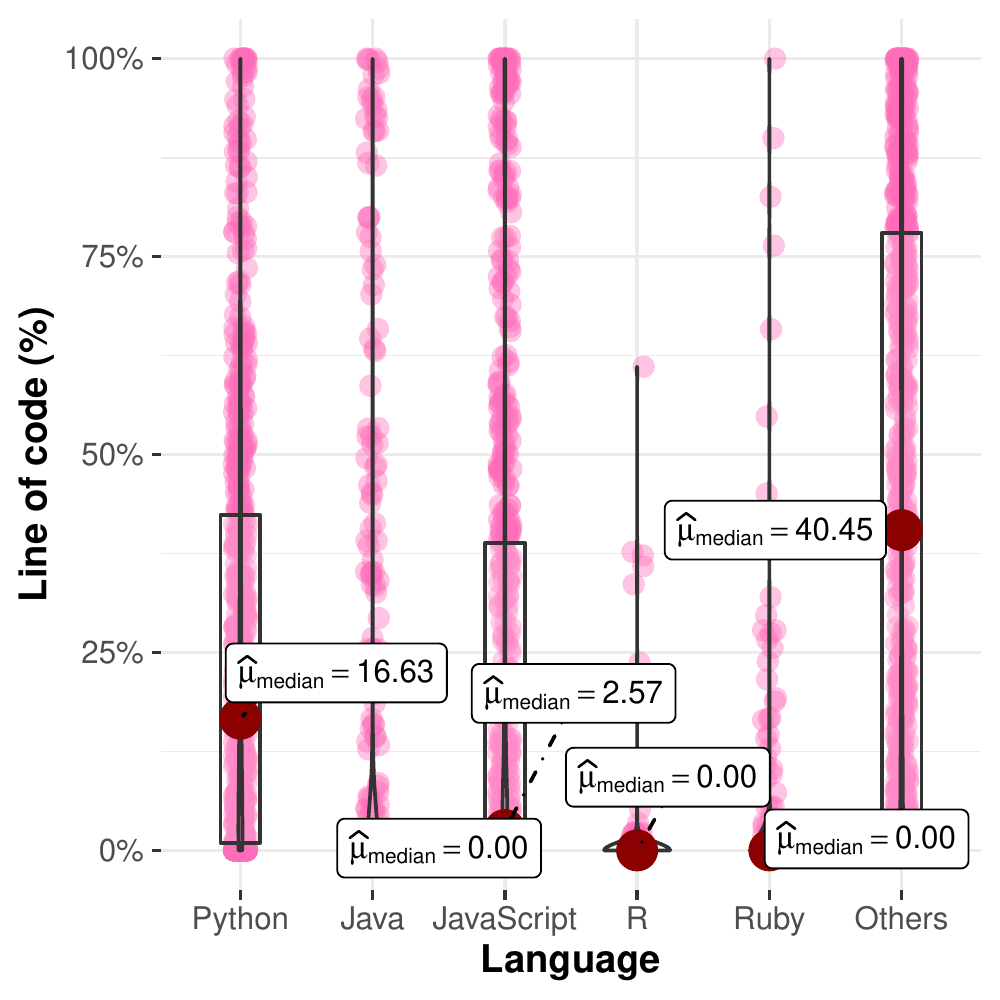}
         \caption{PyPI}
     \end{subfigure}
     \begin{subfigure}{0.3\linewidth}
         \includegraphics[width=\textwidth]{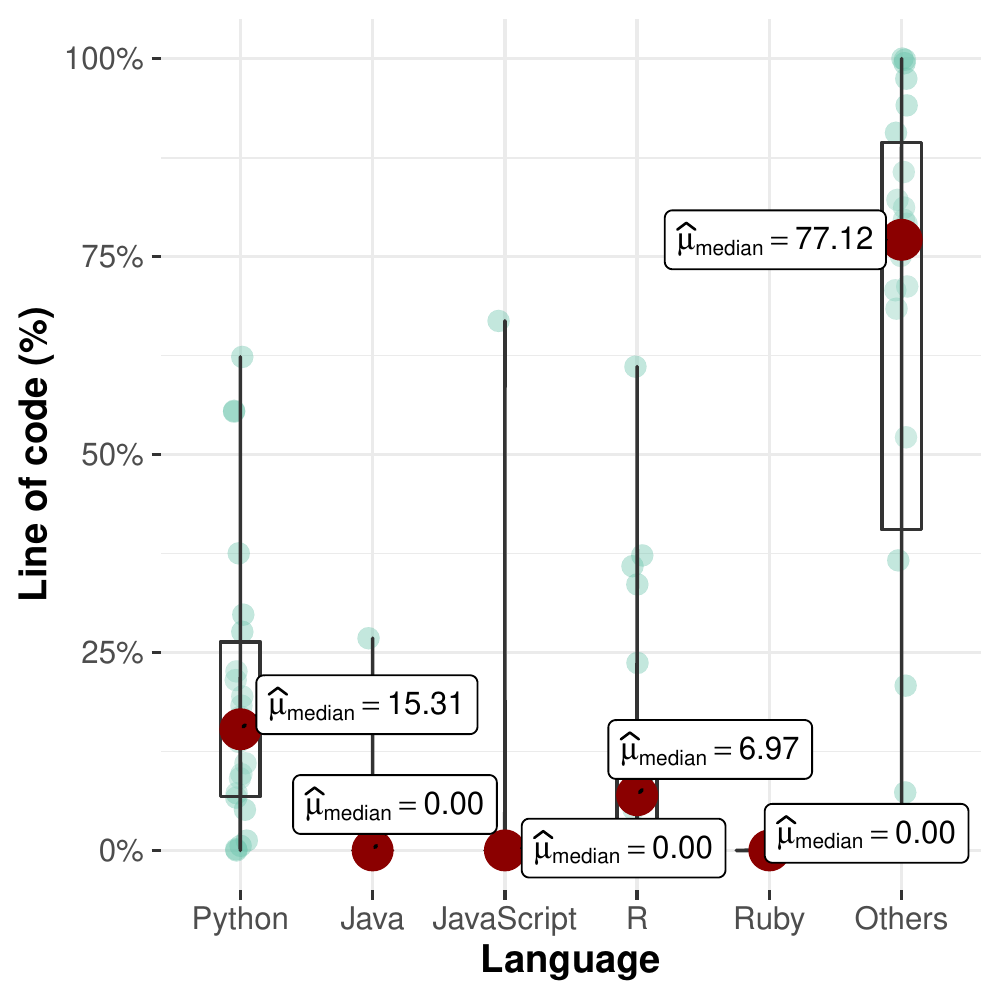}
         \caption{CRAN}
     \end{subfigure}
     \begin{subfigure}{0.3\linewidth}
         \includegraphics[width=\textwidth]{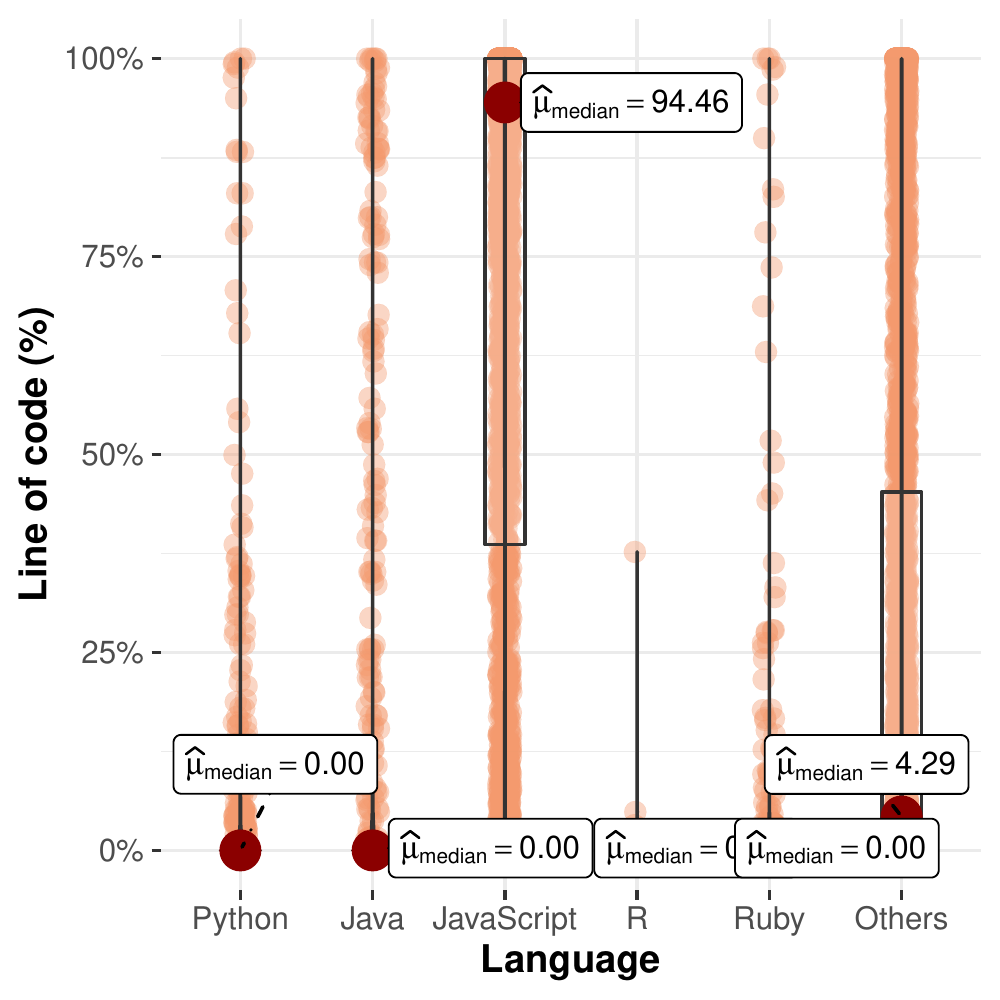}
         \caption{Maven}
     \end{subfigure}
     \begin{subfigure}{0.3\linewidth}
         \includegraphics[width=\textwidth]{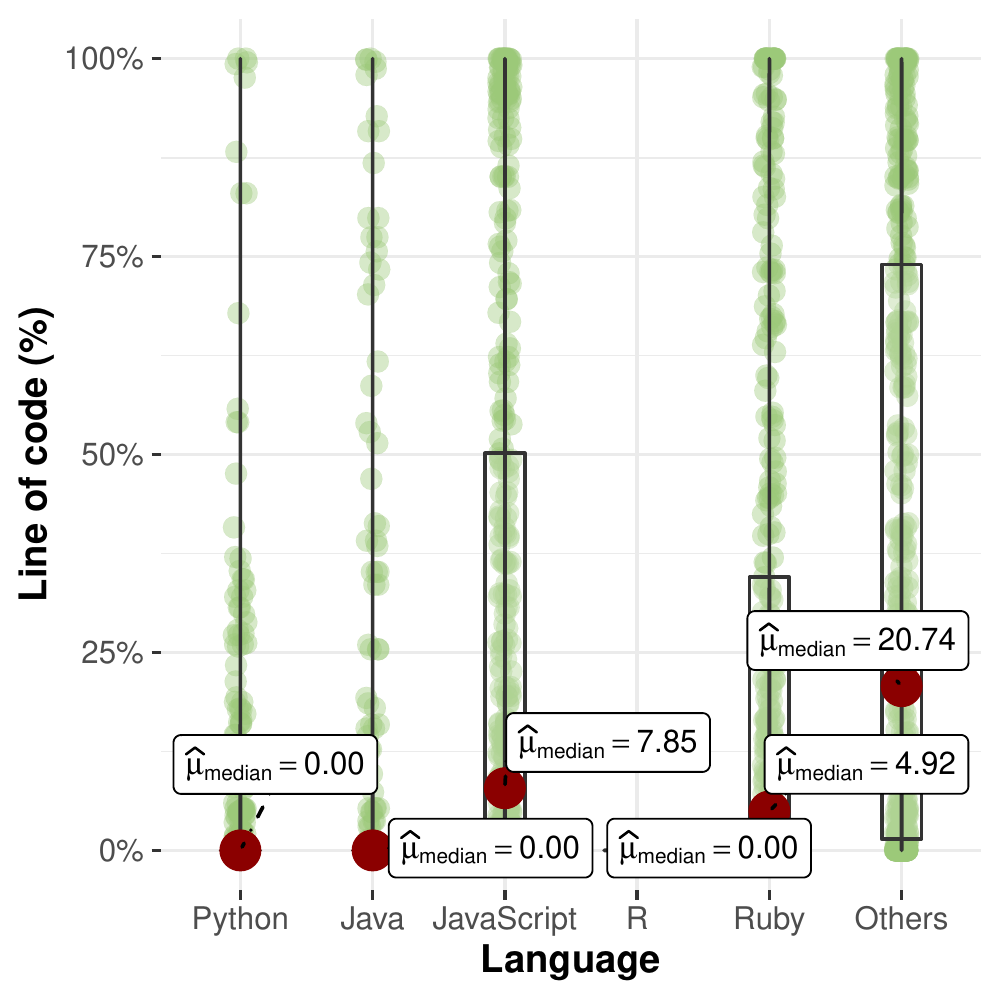}
         \caption{RubyGems}
     \end{subfigure}
    \caption{Answering RQ2, we show the percentage of Programming language per ecosystem.}
    \label{fig:RQ2_result}
\end{figure*}
%%%%%%%%%%%%%%%%%%%%%%%%%%%%%%%%%%%%%%%%%%%%%%%%

In terms of statistical significance, Table~\ref{tab:sta_RQ1} shows evidence that the null hypothesis on whether ‘percentage of contributions from three types are the same’ is rejected. 
We see significant differences (p-value $<$ 0.05) where a library that has a higher percentage of single ecosystem contributors will have a lower percentage of contributors from both ecosystems.
Furthermore, libraries with a higher percentage of independent contributors will have fewer contributors from both ecosystems.

\summary{\RqOneR
}

\subsection{Programming Language Diversity (RQ2)}
\label{sec:approach_rq2}

\paragraph{Approach}
To answer RQ2, we collect and analyze the proportion of programming languages used in a \Cross.
Using our running example, we would like to understand the extent to which the programming language implemented to correspond to the software ecosystem. 
For example, we expect a PyPI library like Bokeh should have a higher proportion of its code written in Python. 
For our analysis, we compare the percentage of the programming language against the five programming languages that correspond with each ecosystem.
This is Python for PyPI, Java for Maven, JavaScript, R for CRAN, and Ruby for RubyGems.
Finally, we grouped all other languages as others. 
Similar to RQ1, we use the McNemar’s Test and Cohen's $d$ to statistically confirm differences in languages.

%%%%%%%%%%%%%%%%%%%%%%%%%%%%%%%%%%%%%%%%%%%%%%%%%
\begin{table}[]
\centering
\caption{A breakdown of the top five languages for the Others category in Figure 4.}
\label{tab:other_language}
\begin{tabular}{lr}
\hline
\textbf{Other}         & \textbf{Percent (\%)} \\ \hline
HTML                             & 63.75                            \\
Shell                            & 23.92                            \\
CSS                              & 10.17                            \\
Makefile                         & 2.15                             \\
C, C++, C\#, PHP, TypeScript & 0.01                             \\ \hline
Total & 100.00                            \\  \hline
\end{tabular}
\end{table}
%%%%%%%%%%%%%%%%%%%%%%%%%%%%%%%%%%%%%%%%%%%%%%%%

\paragraph{\textit{A \Cross~is written using multiple programming languages.}}
Figure~\ref{fig:RQ2_result} shows evidence that a majority of code written for three out of the five ecosystems is from other languages.
Specifically for PyPI it is 40.45\%, 77.12\% for \Cross~that serves the CRAN ecosystem and 20.74\% from RubyGem libraries.
The second related finding is that JavaScript is a primary programming language, especially for NPM (85.01\%), Maven ( 94.45\%), and RubyGem (7.85\%) libraries.
Table \ref{tab:other_language} shows a deeper analysis of the other languages shown in Figure \ref{fig:RQ2_result}. 
We can see that many of the code for a \Cross~is written in HTML, which could be accounted for either documentation and other kinds of meta files that are needed in a library.
This result suggests that a significant portion of a \Cross~is for documentation and other configuration files (Shell, CSS, and makefiles).

Table~\ref{tab:sta_RQ2} shows the statistical results, where the null hypothesis of whether ‘\Cross~written in corresponding programming languages’ is rejected.
We find each ecosystem is significant and mostly uses other languages besides programming languages that serve the ecosystem. 
Additionally, we find that NPM and Maven have significantly used JavaScript more than other languages.

\summary{
\RqTwoR }

\begin{table}[]
\centering
\caption{Contingency Table that shows statistical significance test results that relate to RQ2.}
\label{tab:sta_RQ2}
\scalebox{0.9}{
\begin{tabular}{cccccc}
\cline{2-6}
\multicolumn{1}{l|}{}                   & \multicolumn{5}{c|}{\textbf{Lines of Code (\%)}}                                                                                                      \\ \cline{2-6} 
\multicolumn{1}{l|}{}                   & \textbf{Python}           & \textbf{Java}             & \textbf{JavaScript}       & \textbf{R}                & \multicolumn{1}{c|}{\textbf{Ruby}}             \\ \hline
\multicolumn{1}{|c|}{\textbf{NPM}}      & * L                       & * L                       & * N                       & * L                       & \multicolumn{1}{c|}{* L}                       \\
\multicolumn{1}{|c|}{\textbf{PyPI}}     & * M                       & * L                       & * M                       & * L                       & \multicolumn{1}{c|}{* L}                       \\
\multicolumn{1}{|c|}{\textbf{CRAN}}     & * L                       & * L                       & - & * L                       & \multicolumn{1}{c|}{*L}                        \\
\multicolumn{1}{|c|}{\textbf{Maven}}    & - & - & * N                       & - & \multicolumn{1}{c|}{-} \\
\multicolumn{1}{|c|}{\textbf{RubyGems}} & * L                       & * L                       & * S                       & * L                       & \multicolumn{1}{c|}{* S}                       \\ \hline
\multicolumn{6}{l}{\begin{tabular}[c]{@{}l@{}}The effect sizes level: large(L), medium(M), small(S) and negligible(N)\\ *:p-value \textless 0.05\end{tabular}}                                          
\end{tabular}}
\end{table}

%%%%%%%%%%%%%%%%%%%%%%%%%%%%%%%%%%%%%%%%%%%%%%

\section{Implications}
\label{sec:recommendations}
Based on our findings, we now discuss implications and present possible research directions for each stakeholder.

\paragraph{For Library Maintainers} \textit{Should maintainers consider releasing to different ecosystems?}
The good news that is highlighted in the preliminary study is that cross-ecosystem libraries are already relied upon by the different ecosystems.
For maintainers that are considering to release to multiple ecosystems, they should ensure that already have an existing community of contributors that continue to support them. 
This is because RQ1 indicates that most cross-ecosystem libraries are not likely to receive contributions from both ecosystems.
On the other hand, another interesting result in RQ1 was that a \Cross~has contributions that are independent of the ecosystem.
In terms of programming language, we show that a \Cross~is written in a diverse set of programming languages, sometimes not related to the target ecosystem.
Hence, these maintainers should be competent with multiple programming languages. 
Furthermore, we see that our research opens up new research directions, as we are not sure of the motivations why maintainers decide to support multiple ecosystems, and whether or not they see this phenomenon as being beneficial in terms of solving the problem by replacement libraries, and attracting both new contributors or users. 
We consider all these are future work.

\paragraph{For Library Users and Contributors} \textit{Will cross-ecosystem libraries solve the need to find replacement libraries?}
From the results of the preliminary study, we can conclude that these libraries that decided to release to multiple ecosystems have an already established user base of clients in the original ecosystem. 
This result might suggest that the libraries are quality candidates for replacement, and maybe users are more comfortable using a library that comes from the same maintainers of a library that they are already familiar with.
In terms of RQ1, since the results show that these libraries still rely heavily on contributions from the original ecosystem, contributors might consider supporting these libraries, this might also include learning a new programming language, as shown in RQ2.
The results do suggest that with the increase in technology stacks and languages, contributors, users and maintainers may need to become proficient in multiple programming languages.
However, our results do not report that clients are indeed choosing a cross-ecosystem library, so this would have to be future work.
Another interesting avenue is how these libraries will compete with already established libraries in the new ecosystem that provide the same functionalities.
    
\paragraph{For Software Ecosystems and Researchers} \textit{How will cross-ecosystem libraries impact ecosystem-level topics like governance, and management?}.
From a research perspective, the growing intertwining between different ecosystems will bring forth interesting implications at the ecosystem level.
For instance, to what extent do these libraries abide by the specific rules, and regulations that are enforced by each ecosystem? and how are bug fixes and specifically security vulnerabilities propagated through different ecosystems?
Since RQ1 states that their libraries are more likely to receive contributions from the original ecosystem, does that mean that the boundaries between these two ecosystems become closer? Also, RQ2 states that since these libraries are composed of different programming languages, does this mean that software ecosystems do not need to be defined by programming language, and how do very specific communities like the 'Pythonic' culture of the PyPI community react to such statements?
All these questions should have implications and open up new potential research directions for the software engineering community.

\section{Threats to Validity}
\label{sec:threats_to_validity}
In this section, we outline the threats to the validity of our study.

\textbf{Construct Validity -} 
A key threat in the construct validity exists in the matching approach using the GitHub URL \Cross.
Although we do miss out on other cross-libraries that may exist between two ecosystems, except for CRAN, the GitHub platform is the most central repository as evident by their popularity for PyPI, Maven, and RubyGems.
Also, due to this heuristic, we only are able to detect cross-ecosystem libraries hosted on GitHub, which might be a bias against ecosystems that might not share their projects on GitHub.
Since GitHub is one of the largest platform of Open Source Software, especially for the NPM ecosystem, we are confident of our results.

\textbf{Internal Validity -} 
We discuss two threats to internal validity. 
The first threat is in regard to the tool selection for the statistical analysis.
By using different tools and techniques, results may be different, e.g., statistical testing. 
However, we are confident that the threat is minimum since all employed tools and techniques are used in prior works and are well-known to the software engineering community.
We are also aware that our results are dependent on the dataset. 
However, by using two different reliable data sources (libraries.io, and GitHub API), we are confident in our results.

\textbf{External Validity -} 
The results of our analyses are limited to the five ecosystems: NPM, PyPI, CRAN, Maven, and RubyGems, which threaten the generalization of our claims to other ecosystems.
Still, we believe that our analyses could easily be applied to other ecosystems and are seen as immediate future work.

\section{Related work}
\label{sec:related_work}
In this section, we discuss key ideas from related works and how our work aligns and complements the existing literature.

\textbf{On Analogical Libraries and Library Recommendation}
These studies relate to the analogical libraries from various ecosystems.
\citet{Cossette2021recommendation}, highlighted that analogical library recommendation tools should beware of false predictions as the consequences are serious.
\citet{Ouni:2017} proposed an approach to recommend libraries by using a search base algorithm.
They suggested that the history of library usage and semantics of each library can be used to improve the efficiency of the recommendation technique.
\citet{Chen2016QA} presented a new approach to recommend analogical libraries from tags of millions of Stack Overflow questions by using the word embedding technique.
From the study by \citet{6385124}, they identified sets of analogical libraries by using library migration graphs that show how existing libraries have performed migrations among them.
These migration graphs allow developers to discover and select library replacements easier.
Studies such as \citep{nguyen2020crossrec, Thung:WCRE2013} created a  library recommendation system in order to developers capture unaware suitable libraries based on the set of used dependencies in the project.
\citet{chen2019s} created a knowledge base of analogical libraries which mined from tags of millions of Stack Overflow questions.
They then implemented a proof-of-concept web application to recommend an analogical library.
Several studies focused on helping developers to migrate the libraries and API of their project to new ecosystems.
\citet{zhong2010mining} presented a tool that maps API from Java to C\# called MAM (i.e., Mining API Mapping). 
This tool can reduce the errors (e.g., compilation error, defect) during the library migration to the new ecosystem.
\citet{Teyton2013map} automatically extracted function mappings from the analysis of software libraries that previously performed the library migration. 
\citet{Camilo2020MUTAMA} proposes MUTAMA, a multi-label classification approach to the Maven library tagging problem based on information extracted from the byte code of each library. 
In our work, we perform an empirical study on the cross-ecosystem library which is a library that is cross-cutting different library ecosystems.
Prior works, instead, focus on libraries from the same ecosystem.

\textbf{On Software Categorization}
These studies focus on categorizing libraries by using various kinds of their characteristics.
\citet{kawaguchi2006mudablue} proposed MUDABlue, a tool that automatically categorizes software systems by using the information from source code.
MUDABlue allows a software system to be a member of multiple categories.
MUDABlue has three major aspects: (1) it relies on no other information than the source code, (2) it determines category sets automatically, and (3) it allows a software system to be a member of multiple categories.
MUDABlue not only correctly determined the categories as specified by the authors of the software, it determined some additional categories that were missed by the manual categorization. 
We have also shown that MUDABlue is more advanced than the research systems described in the literature.

In this work, we a \Cross~is a kind of software category that spans across ecosystems.
Hence, it would be interesting to understand how \Crosses~are categorized.

\textbf{On Open Source Software Retention, Engagement, and Sustained Contributions} - The common motivations to make contributions to OSS projects are the joy of programming, the identification with a community, career advancement, and learning \citep{Hars:2001}.
Additionally, \citet{Roberts:2006} explored the interrelationships between OSS developer motivations, revealing the motivations are not always complementary.
When comparing motivations between individual developers and companies, \citet{Bonaccorsi:2006} observed that companies are more motivated by economic and technological reasons.

In this work, we would like to understand how cross-ecosystem libraries are able to engage and retain contributions from multiple ecosystems.
In this case, the motivation is to ensure that the library can have sustained activity, especially if the ecosystem becomes dependent on this library.

\section{Conclusion and Future Directions}
\label{sec:conclusion}
Developers may inevitably face the need to switch programming languages, which ultimately may result in replacing third-party libraries.
An alternative to replacement libraries is for maintainers to release different releases for different ecosystems (i.e., \Cross).
In a large-scale study that covers over 1.1 million libraries, we find that contributors to a \Cross~belong to one ecosystem, and more likely to be implemented using multiple programming languages.
Insights from this study lay the groundwork and open up new opportunities for library maintainers, users, contributors, and researchers into understanding how these different ecosystems are becoming more intertwined with each other.

% \section{Acknowledgement}
% This work was supported by JSPS KAKENHI Grant JP20H05706.

\section*{Declaration of Interest}

\textbf{Funding-} This work has been supported by JSPS KAKENHI Grant Number JP20H05706, JP20K19774\\

\textbf{Conflicts of interest-} Raula Gaikovina Kula is on the Editorial Board.\\

\textbf{Availability of replication package-} Our replication package is openly available at \url{https://zenodo.org/record/6983864#.YvVmGuxBy3I}.\\

\textbf{Author contributions-} 
\begin{itemize}
    \item \textit{Kanchanok Kannee}: Conceptualisation, Methodology, Software, Original Writing draft, Visualisation.
    \item \textit{Supatsara Wattanakriengkrai}:  Investigation, Data collection, Original Writing, Review.
    \item \textit{Ruksit Rojpaisarnkit}: Data Curation, Investigation, Original Writing
    \item \textit{Raula Gaikovina Kula}: Conceptualisation, Funding Acquisition, review and editing drafts, Supervision, project administration.
    \item \textit{Kenichi Matsumoto} : Funding Acquisition, review and editing drafts, Supervision, project administration.
\end{itemize}

\bibliographystyle{elsarticle-num-names}
\bibliography{bibliography}

\end{sloppy}
\end{document}
\endinput